\documentclass[aps,pre,amsmath,amsfonts,floatfix,superscriptaddress]{revtex4}

\usepackage{graphicx}

\font\msytw=msbm9 scaled\magstep1

\let\a=\alpha     \let\d=\delta \let\e=\varepsilon
\let\z=\zeta     \let\th=\theta  
             \let\p=\pi    \let\r=\rho
\let\s=\sigma    \let\f=\varphi 
   
\let\G=\Gamma

\let\io=\infty
\def\ie{{i.e. }}\def\eg{{e.g. }}

\def\FF{{\cal F}}

\def\ZZ{{\cal Z}}

\def\Re{{\rm Re}\,}\def\Im{{\rm Im}\,}

\def\to{\rightarrow}
\def\la{\left\langle}
\def\ra{\right\rangle}

\def\RRR{\hbox{\msytw R}}

\newcommand{\beq}{\begin{equation}}
\newcommand{\eeq}{\end{equation}}
\newcommand{\wh}{\widehat}
\newcommand{\wt}{\widetilde}
\newcommand{\Tr}{\text{Tr}}

\begin{document}

\title{On the high density behavior of Hamming codes with fixed minimum distance}

\author{Giorgio Parisi}

\affiliation{Dipartimento di Fisica, Universit\`a di Roma ``La Sapienza'', 
P.le A. Moro 2, 00185 Roma, Italy}

\affiliation{INFM -- CRS SMC, INFN, Universit\`a di Roma ``La Sapienza'', 
P.le A. Moro 2, 00185 Roma, Italy}

\author{Francesco Zamponi}

\affiliation{Laboratoire de Physique Th\'eorique de l'\'Ecole Normale Sup\'erieure,
24 Rue Lhomond, 75231 Paris Cedex 05, France}

\date{\today}

\begin{abstract} 
We discuss the high density behavior of a system of hard spheres of diameter $d$ on the
hypercubic lattice of dimension $n$, in the limit $n\to\io$, $d\to\io$,
$d/n=\d$. 
The problem is relevant for coding theory, and the 
best available bounds state that the maximum density of the system
falls in the interval $1 \leq \r V_d \leq \exp(n \kappa(\d))$, being
$\kappa(\d) > 0$ and $V_d$ the volume of a sphere of radius $d$. 
We find a solution of the equations describing the liquid
up to an exponentially large value of $\wt \r = \r V_d$, but we show that this
solution gives a negative entropy for the liquid phase for $\wt \r \gtrsim n$. We then
conjecture that a phase transition towards a different phase might take place,
and we discuss possible scenarios for this transition.
\end{abstract} 


\maketitle

A code $C$ is a subset of the binary Hamming space $H_2^n=\{0,1\}^n$. The distance between
two points $s,t \in H_2^n$ is the Hamming distance 
$d(s,t) = \sum_{\a=1}^n (s_\a + t_\a)_{\text{mod}2}$, \ie it is given by the number of different
bits.
We consider the problem
of finding the maximal size $A(n,d)$ of a code $C$ such that the minimum distance between two
points in $C$ is $d$, that means, denoting by $|C|$ the number of sequences in $C$,
\beq
A(n,d) = \max \Big[ \, |C| \, \Big| \, \forall s,t \in C , \, d(s,t) \geq d \Big] \ .
\eeq
In particular we are interested in the quantity
\beq\label{R}
R(\delta) = \limsup_{n\to\io, \, d/n \to \d} \frac{1}{n}\log_2 A(n,d) \ ,
\eeq
where the supremum is taken on all possible sequences of codes such that $d/n \to \d$.
The problem trivializes for $\d > 1/2$ as the total number of sequences is finite and 
$R(\delta)=0$. An interesting scaling is $d/n = 1/2 - \e n^{-\a}$, as for an appropriate choice of
$\a$ the number $A(n,d)$ might increase polynomially in $n$, but this will not be investigated
here. Thus we will restrict to $\d < 1/2$ in the following.

This problem is relevant for the theory of error correcting codes 
\cite{VG,PS99,MRRW77,BJ01,Sa01,De73}. In ``physics language'', it is the problem of
finding the maximum possible density of a system
of hard spheres on the hypercubic lattice.
This rephrasing of the problem has been shown to be useful as it allows to
use well known methods borrowed from the theory of liquids, like
the virial expansion~\cite{PS99}.

In this paper we will discuss the behavior of the system at high density, in
order to understand how one can try to compute the maximum density.
We will show that, for large $n$, the problem closely resembles the problem of hard
spheres in $\RRR^n$ in the limit of large space dimension $n$. The basic idea
is that in this limit the number of neighbors of a sphere is large, much as it happens in the
continuum for large space dimension.
We will then discuss some recent ideas that have been used in the continuum
\cite{FP99,PS00,PZ06,To06} to make some progress in the direction of deriving 
bounds on $A(n,d)$.

The best known lower bound on $A(n,d)$ (Varshamov-Gilbert bound) states that the density
$\r = |C|/2^n \geq 1/V_{d-1}$, $V_d$ being the volume of a sphere of radius $d$ in 
$H_2^n$~\cite{VG}.
This bound can be proven from the convergence of the virial series~\cite{PS99} and
gives $R(\d) \geq R_{VG}(\d) = 1- H(\d)$, $H(\d)$ being the binary entropy function (see below).
This means that a ``liquid phase'', defined by the virial equation of state, exists at least
up to $\r V_{d-1} \sim 1$. We will show that the liquid phase can be formally continued up to a density
$\r V_{d-1} \sim \exp(n \kappa(\d))$ with $\kappa(\d) > 0$: this result correspond to the Frisch-Percus
result in the continuum \cite{FP99}.

However, we find that the entropy of the liquid becomes negative at $\r V_{d-1} \sim n$. This suggests
the possible instability of the liquid towards a different phase, \ie the
existence of a {\it phase transition}. We will discuss two different
possibilites. By analogy with the problem in $\RRR^n$ for large $n$, we
conjecture that a glass transition might be present also in this system.
In absence of other phases
(such as ``crystalline'' phases), this analogy suggests that $R(\d)$ is given by the Varshamov-Gilbert
result, $R(\d) = R_{VG}(\d)$.
We will also discuss a different instability that happens for even $d$, leading to a first order
transition to a phase where the particles move on a sublattice of $H^n_2$.

The calculation of the properties of the (eventual) glassy phase requires the
use of the replica method, but this turns out to be more difficult than in
$\RRR^n$ so we leave it for future work.

The paper is organized as follows: in section~\ref{sec:definitions} we set up
the basic notations and definitions; in section~\ref{sec:bounds} we review the
known bounds on $R(\d)$; in section~\ref{sec:bulk} we present our main results
and conjectures on the high density behavior of the system; finally in
section~\ref{sec:conclusions} we draw the conclusions and present a summary of
our ideas.

\section{Hard spheres on the hypercube}
\label{sec:definitions}

\subsection{Definitions}

Let $s = (s_1,\cdots,s_n)$, $s_\a \in \{0,1\}$, be a point in $H^n_2$.
We define the Hamming distance between two points
\beq\label{Hamdist}
d(s,t)= \sum_{\a=1}^n (s_\a+t_\a)_{\text{mod}2} \ .
\eeq
The number of
points at distance $i$ from the origin is $S_i \equiv \binom{n}{i}$ and the 
volume of a sphere of radius $d$ centered in the origin is
\beq
V_d \equiv \sum_{t : d(0,t)\leq d} 1 =  \sum_{i=0}^d \binom{n}{i} \ .
\eeq
The interaction potential between two particles in $s$ and $t$ is the {\it hard core} potential:
\beq
U(s,t) = \begin{cases} 0 \hskip15pt d(s,t)\geq d \ , \\
\infty \hskip10pt d(s,t) < d \ . \end{cases}
\eeq
The grancanonical partition function \cite{Hansen,PS99} is
\beq\label{GCpart}
\ZZ_n[z] = \sum_{N=0}^{2^n} \frac{z^N}{N!} \sum_{s_1\cdots s_N} e^{-\sum_{i<j} U(s_i,s_j)} \ ,
\eeq
the average number of spheres (to be identified with the average size of the code, $|C|$)
is $\la N \ra = z\frac{\partial \ln \ZZ_n}{\partial z}$ and
$A(n,d) = \lim_{z\to\io} \la N \ra$. The density is $\rho \equiv \la N \ra/2^n$ and it is convenient
to define the {\it reduced density} 
\beq
\wt \rho \equiv \rho V_{d-1} = \frac{\la N \ra V_{d-1}}{2^n} \ ,
\eeq
and
\beq\label{fdefi}
\f = \frac1n \ln \wt \rho \ .
\eeq
For $n\to\io$ and $d/n=\d < 1/2$ one has
\beq
v(\d) = \lim_{n\to\io} \frac1n \ln V_{n\d} = -\d \ln\d - (1-\d)\ln(1-\d) \ ,
\eeq
\ie $V_d \sim S_d = \binom{n}{d} \sim e^{nv(\d)}$, so that
\beq\label{fdef}
\f = \frac1n \ln \wt \r \sim \frac1n \ln \r + v(\d) \ , \hskip10pt
\r \sim e^{n [\f - v(\d)]} \ .
\eeq
The pair distribution function is given by the average number of particle pairs such
that one particle is in $s$ and the other in $t$~\cite{Hansen}:
\beq\label{gdef}
\r_2(s,t) \equiv \la \sum_{i\neq j} \d_{s,s_i} \d_{t,s_j} \ra \equiv \r^2 g(d(s,t))
\eeq
where the Kronecker $\d_{s,t}$ is equal to one if $s=t$ and to zero otherwise.
The pair distribution function $g(i)$ is normalized to have value $1$ at large distance
(for $n\to\io$). We also define $h(i) \equiv g(i)-1$ which vanishes at large
distance.
Note that the function $\r g(d(0,s))$ is the probability of finding a particle
in the point $s$, given that there is a particle in the origin.

\subsection{Liquid phase}

The liquid phase is defined by the virial expansion of the partition function at low
densities which converges for $\r V_{d-1} \leq 1/6e$ \cite{PS99}.
However, one can try to perform a continuation of the liquid
equation of state below the radius of convergence of the virial series. 
For hard spheres in $\RRR^3$, such a continuation is possible:
approximated expressions are usually obtained by resumming some class of diagrams of
the virial expansion to get a closed integral equation for the pair
correlation function $g(i)$. Well known resummations are the Percus-Yevick
(PY) and the HyperNetted Chain (HNC) ones, and agree well 
with numerical results \cite{Hansen}.

For hard spheres in $\RRR^n$ it has been shown by Frisch and Percus
\cite{FP99} that in the limit of large $n$ the virial series is dominated
order by order by the so-called {\it ring diagrams}.
The resummation of the ring diagrams has been shown to provide a reasonable analytic 
continuation of the liquid equation of state up to very large values of the density \cite{FP99}.
As the latter diagrams are included in the HNC resummation, the two resummations should be
equivalent in the large dimension limit. The advantage of the HNC resummation is that it
leads to a closed expression for the free energy corresponding to the partition function
(\ref{GCpart}) for any finite $n$. The HNC free energy (per particle) is:
\beq
\label{HNCFree}
\Psi[g] = \frac{\r}{2} \sum_{i=0}^n S_i \big[ g(i) \ln g(i) - g(i) + 1 \big]
+ \ln \r -1 + \frac{1}{2 \, \r \, 2^n} \sum_{p\geq 3} \frac{(-1)^p \r^p}{p} \Tr h^p \ ,
\eeq
where
\beq
\Tr h^p = \sum_{s_1 \cdots s_p} h(s_1,s_2) h(s_2,s_3) \cdots h(s_p,s_1) \ ,
\eeq 
and the function $g(i)$ is determined by the stationarity condition $\frac{\d \Psi}{\d g}=0$
({\it HNC equation}).
We will argue that a solution to the HNC equations (or to similar approximations to the
equation of state of the liquid, that should be equivalent for $n\to\io$)
exists up to $\wt\r \sim \exp(n\kappa)$ with
$\kappa > 0$, thus defining the continuation of the liquid equation of state in this
region of densities.

\subsection{Fourier transform}

To write the last term of the HNC free energy in a more convenient way it is useful to define the 
Fourier transform
on the hypercube. Define the scalar product of two sequences as 
$s \cdot t \equiv \sum_{\a = 1}^n s_\a t_\a$ (such that $s^2 \equiv s\cdot s = d(0,s)$) 
and let us indicate by $s+t$ the sum (modulo 2)
of the two sequences. 
Then if $q \in \{0,1\}^n$ the Fourier transform of a function $h(s)$ 
is given by
\beq\label{FT}
\begin{split}
\wh h(q) &= \sum_s (-1)^{q\cdot s} h(s) \ , \\
h(s) &= \frac1{2^n} \sum_q (-1)^{q\cdot s} \wh h(q) \ ,
\end{split}
\eeq
Using the property $\sum_s (-1)^{(q_1 + q_2) \cdot s} = 2^n \d_{q_1,q_2}$ the Fourier transform
of a function $h(s,t) = h(s+t)$ is
\beq\label{FTtrasl}
\begin{split}
&\sum_{s,t} (-1)^{q_1 \cdot s} (-1)^{q_2 \cdot t} h(s+t) = 2^n \d_{q_1,q_2} \wh h(q_1) \ , \\
&h(s,t) = \frac{1}{2^n} \sum_q (-1)^{q \cdot (s+t)} \wh h(q) \ .
\end{split}
\eeq
Moreover if $h(s,t) = h(d(s,t))$ (a {\it rotationally invariant} function), 
its Fourier transform depends only on $a \equiv q^2 = d(0,q)$, see Eq.~(\ref{Hamdist}), and one has
\beq\label{FTsferico}
\begin{split}
&\wh h(a) = \sum_{i=0}^n \FF_n(a,i) h(i) \ , \\
&h(i) = \frac{1}{2^n} \sum_{a=0}^n \FF_n(i,a) \wh h(a) \ , \\
&\FF_n(a,i) = \sum_{s : d(0,s)=i} (-1)^{\sum_{\a=1}^a s_\a} =
\sum_{m=\max(0,i+a-n)}^{\min(a,i)} (-1)^m \binom{a}{m} \binom{n-a}{i-m} = K_i(n,a) \ ,
\end{split}
\eeq
where $K_i(n,a)$ is a {\it Krawtchouk polynomial} \cite{Sa01,kraw}.
The matrix $\FF_n(a,i)$ has the following symmetries that will be useful in the following:
\begin{eqnarray}
&\binom{n}{a}\FF_n(a,i) = \binom{n}{i} \FF_n(i,a) \label{S1} \ , \\
&\FF_n(n-a,i) = (-1)^i \FF_n(a,i) \label{S2} \ , \\
&\FF_n(a,n-i) = (-1)^a \FF_n(a,i) \label{S3} \ , \\
&\FF_n(n-a,n-i) = (-1)^{n-a-i} \FF_n(a,i) \label{S4} \ .
\end{eqnarray}
It satisfies $\FF_n(0,i) = \binom{n}{i}$ and $\FF_n(a,0) = 1$, and can be 
easily constructed using the recursion relations
\beq
\begin{split}
&\FF_{n+1}(a,0) = 1 \ , \\
&\FF_{n+1}(a,i+1)= \FF_n(a,i) + \FF_n(a,i+1) \ , \hskip20pt a \leq n \ , \\
&\FF_{n+1}(n+1,i+1) = -\FF_n(n,i) + \FF_n(n,i+1) \ . 
\end{split}
\eeq
Using Eq.~(\ref{FTtrasl}) it is easy to show
that
\beq
\Tr h^p = \sum_{q} [\wh h(q)]^p \ ,
\eeq
and the HNC free energy (\ref{HNCFree}) becomes, defining $L_3(x) = \ln (1+x) - x + \frac{x^2}2$,
\beq
\label{HNCFree2}
\Psi[g] = \frac{\r}{2} \sum_{i=0}^n S_i \big[ g(i) \ln g(i) - g(i) + 1 \big]
+ \ln \r -1 - \frac{1}{2 \, \r \, 2^n} 
\sum_{a=0}^n S_a L_3[\r \wh h(a)] \ ,
\eeq
and the HNC equation can be written as~\cite{Hansen}
\beq\label{HNCeq}
\begin{split}
&\ln g(s) = h(s) - c(s) \ , \\
&\wh c(q) = \frac{\wh h(q)}{1 + \r \wh h(q)} \ .
\end{split}
\eeq

\section{Known bounds on $R(\d)$}
\label{sec:bounds}

In this section we will review some known bounds on $R(\d)$.
Using Eq.s~(\ref{R}), (\ref{fdef}), and $\r = \la N \ra / 2^n$, 
$R(\d)$ is related to the maximum density $\f_c(\d)$ of the spheres by
\beq
R(\d) = 1 - H(\d) + \frac{\f_c(\d)}{\ln 2} \ ,
\eeq
where $H(\d) = \frac{v(\d)}{\ln 2} = -\d \log_2 \d - (1-\d) \log_2 (1-\d)$ is the
{\it binary entropy function}. We will use units of $\f$ because it will lighten the
notation in section \ref{sec:bulk}.

As discussed in the introduction,
the best lower bound for $R(\d)$ (Varshamov-Gilbert bound \cite{VG}) can be obtained by proving the 
convergence of the virial series \cite{PS99}. It turns out that the virial series converges for 
$\r V_{d-1} = \wt \r < 1/6e$, that means $\f < 0$ for $n \to \io$. Thus $\f_c \geq 0$ and a lower
bound for $R$ is $R(\d) \geq 1 - H(\d)$.

A trivial upper bound follows from the fact that the total volume occupied by the
spheres is $\sim N V_{d/2}$ and should obviously be smaller than the total volume
$2^n$. Thus
\beq\label{fmax}
\r = \frac{N}{2^n} < \frac1{V_{d/2}} \sim e^{-n v(\d/2)}
\hskip10pt \Leftrightarrow
\hskip10pt
\f < \f_{max} \equiv v(\d) - v(\d/2) \ ,
\eeq
so $1-H(\d)+\f_{max}/\ln 2=1-H(\d/2)$ is an upper bound for $R(\d)$.

Better upper bounds for $R(\d)$ can be derived by Delsarte's linear programming method~\cite{De73}.
In physics language, they follow from the observation that the minimal requirements
for the correlation function $g(i)$ are the following: \\
{\it i)} $g(i) = 0$ for $0\leq i < d$, as no pair of particles can be
at a distance smaller than $d$ due to the hard core interaction. \\
{\it ii)} $g(i) \geq 0$, $\forall i$ ; this follows from the definition of $g(i)$, 
see Eq.~(\ref{gdef}). \\
{\it iii)}
$\r \wh h(a) \geq -1$, $\forall a$; this is because the structure factor $S(a)= 1+\r\wh h(a)$ is
a positive quantity, equal to the average of the square modulus of the $q$-component of
density fluctuations~\cite{Hansen}.

If one is able to find a function
$g(i)$ verifying conditions {\it i)-iii)} above (that have sometimes 
been called {\it positivity conditions}, see \eg \cite{To06}) for a given value of $\r$, 
one can compute the corresponding value of $\la N \ra$ either as $\la N \ra=\r 2^n$ or by 
working in the canonical ensemble (\ie at fixed $N$ \cite{Hansen}) and
recalling that from Eq.~(\ref{gdef}) it follows:
\beq\label{enneN}
N ( N-1) = \sum_{s,t} \r^2 g(d(s,t)) = N  \sum_{i=0}^n \binom{n}{i} \r g(i) 
\hskip10pt \Rightarrow \hskip10pt 
N  = 1 + \sum_{i=0}^n \binom{n}{i} \r g(i) \ .
\eeq
It is not obvious that one can find configurations of the system with density $\r$ 
that actually produce the function $g(i)$, see \eg \cite{Le75,To06,Le06} for a discussion of this
issue in the case of spheres in the continuum.
However, as the $\r$, $g(i)$ obtained from the partition function
(\ref{GCpart}) (or from the canonical partition function)
must satisfy conditions {\it i)-iii)}, it is clear
that the value of $A(n,d)$ is smaller than the maximum of the right hand side of Eq.~(\ref{enneN})
over all the possible choices of $\r$, $g(i)$ satisfying the positivity conditions, i.e.
\beq\label{bound1}
A(n,d) \leq \max \left[1 + \sum_{i=0}^n \binom{n}{i} \r g(i) \,
\Big| \, \r , \, g(i) \, : \text{ {\it i)-iii)}}   
\right]
\ .
\eeq
The problem with Delsarte's method is that it can be used to derive arithmetically an upper bound
for finite and not too large $n$ (\eg $n=1000$, see \cite{BJ01}), 
but it is not easy to obtain analytical results for $n\to\io$.

\begin{figure}[t]
\includegraphics[width=10cm]{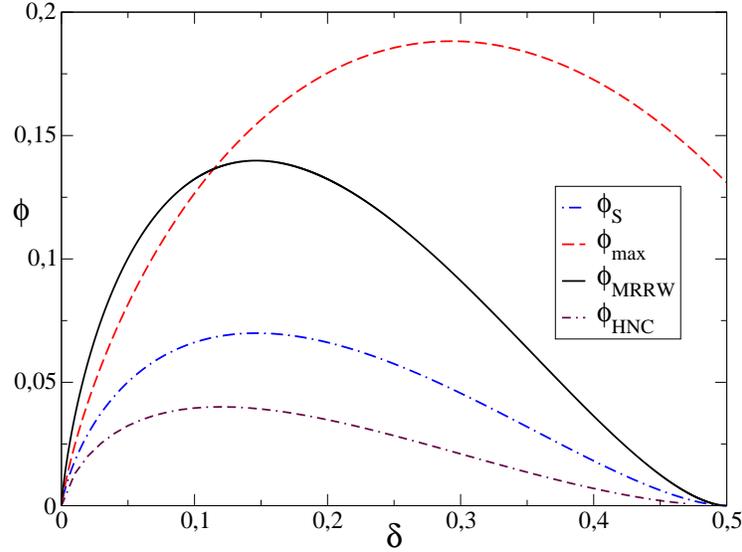}
\caption{Bounds for $\f_c$. The VG lower bound is $\f_c \geq 0$. The upper
bounds $\f_{max}(\d)$ (\ref{fmax}), dashed line, and $\f_{MRRW}(\d)$
(\ref{MRRW}), full line, are reported.
The lower bound to $D(\d)$ obtained by Samorodnitsky, $\f_S(\d)=\f_{MRRW}(\d)/2$, is reported as a dot-dashed
line. Finally, the value of $\f_{HNC}(\d)$, below which our solution for the liquid state is well defined, 
is reported as a dot-dashed-dashed line.
}
\label{fig:1}
\end{figure}

In the literature the
Delsarte method has been often formulated in terms of the function
$A_i =\d_{i0} + \r \binom{n}{i} g(i)$, see \eg \cite{BJ01,Sa01}. Conditions {\it i)-ii)} are equivalent
to $A_0=1$, $A_i=0$ for $1 \leq i < d$ and $A_i \geq 0$. Condition {\it iii)} gives, using 
Eq.s~(\ref{FTsferico}) and (\ref{S1}), and recalling that $\FF_n(a,0)=1$, $\FF_n(0,i)=\binom{n}{i}$ 
and $\sum_{i=0}^n \FF_n(a,i) = 2^n \d_{a0}$ (being the Fourier transform of the function 1):
\beq\label{contoS}
\begin{split}
S(a) &= 1 + \r\wh h(a) = 1 + \sum_{i=0}^n \FF_n(a,i) [\r g(i)- \r] =
1+\sum_{i=0}^n \frac{ \FF_n(i,a)}{\binom{n}{a}}\binom{n}{i} \r  g(i) - \r 2^n \d_{a0} \\ &=
\binom{n}{a}^{-1} \sum_{i=0}^n \FF_n(i,a) \left[\r\binom{n}{i} g(i) + \d_{0i}\right] - \r 2^n \d_{a0}
= \binom{n}{a}^{-1}\sum_{i=0}^n \FF_n(i,a) A_i - \r 2^n \d_{a0} \ .
\end{split}
\eeq
Thus the condition $S(a) \geq 0$ is equivalent to $\sum_{i=0}^n \FF_n(i,a) A_i \geq 0$ for $a \neq 0$,
while for $a=0$ we simply have $\sum_{i=0}^n A_i = N$ from Eq.~(\ref{enneN}).
Then the bound (\ref{bound1}) can be reformulated as 
\beq\label{Delsarte}
A(n,d) \leq \max \left[ \, \sum_{i=0}^n A_i \, \left| \, A_i \geq 0 \, \,\, \forall i \, ;
A_0=1 \, ; A_1=\cdots=A_{d-1}=0 \, ; \sum_{i=0}^n \FF_n(i,a) A_i \geq 0 \,\,\, \forall a \right. \right]
\equiv D(n,d) \ .
\eeq
Upper bounds for $A(n,d)$ can be derived by studying the dual to the linear problem (\ref{Delsarte}), 
see \eg \cite{BJ01}.
In this way one can prove that
(MRRW bound \cite{MRRW77}), for $n\to\io$:
\beq
D(\d) \equiv \limsup_{n\to\io , n/d\to \d} \frac1n \log_2 D(n,d)
 \leq H\left(\frac12 - \sqrt{\d(1-\d)}\right) \equiv R_{MRRW}(\d) \ ,
\eeq
that means that $R(\d)$ verifies the same bound and 
\beq\label{MRRW}
\f \leq v\left(\frac12- \sqrt{\d(1-\d)}\right)-\ln 2 + v(\d) \equiv \f_{MRRW}(\d) \ .
\eeq
This bound has been (little) improved only for $\d < 0.273$ \cite{MRRW77,BJ01}.

By constructing an explicit solution to the positivity conditions, one can obtain lower
bounds on $D(\d)$ (but this is not sufficient to have the same bound for $R(\d)$, \cite{To06,Le06}). 
For this problem, this has been done in \cite{Sa01}, where it was shown that
\beq\label{gdelta}
g(i) = (d+1) \d_{id} + \th(i-d) \ ,
\eeq
where $\th(i-d) = 1$ for $i>d$ and $0$ otherwise, verifies conditions {\it i)-iii)} up
to a density $\f_S(\d) = \f_{MRRW}(\d)/2$.

To resume, the situation is the following:
\begin{itemize}
\item the VG bound states that the maximum density $\f_c \geq 0$;
\item the MRRW bounds coming from the dual to (\ref{Delsarte}) prove that $\f_c < \f_{MRRW}(\d)$,
Eq.~(\ref{MRRW}) (with a little improvement for $\d < 0.273$, see \cite{BJ01});
\item the Delsarte's method cannot be used to prove that the VG bound is tight, \ie that $\f_c \leq 0$, 
as the lower bound on
$D(\d)$ obtained in \cite{Sa01} implies that the best upper bound from Delsarte's method cannot be
smaller than $\f_{MRRW}(\d)/2$;
\item moreover, recent arithmetical results for $n=1000$ \cite{BJ01}
seems to indicate that the actual value of $D(\d)$ might be close to the MRRW ones, \ie much larger
then the lower bound of \cite{Sa01}. This means that the MRRW bounds are
probably very close to the best one
can obtain from the Delsarte's method (\ref{Delsarte}).
\end{itemize}
The bounds above for $\f_c$ (summarized in Fig.~\ref{fig:1})
leave a large gap - at least of the order of $\f_{MRRW}(\d)/2$, but probably
of the order of $\f_{MRRW}(\d)$ - and there are not so many ideas on how to improve 
them~\cite{BJ01}. Note that, as discussed in the introduction, for $\d > 1/2$ one can
prove that $R(\d)=0$, so the size of the code is not exponential.

\section{The liquid phase at high density}
\label{sec:bulk}

In this section we will discuss some insight on the problem that comes from
the physical intuition on the possible behavior of the system (\ref{GCpart}). We are
not able to present rigorous results but we
hope that the discussion below 
will lead to new ideas on how to rigorously improve the bounds on $R(\d)$.

It is convenient to outline our basic ideas before going into the details of the calculations.
We try to find a solution to the HNC equations (\ref{HNCFree}), (\ref{HNCeq}) 
(or to other approximate equations for the liquid) for $\f > 0$ ($\f$ is defined in Eq.~(\ref{fdefi})). 
We assume here that any resummation will be equivalent for $n\to\io$ as long
as it includes the ring diagrams \cite{FP99}.
Such a solution should clearly verify {\it at least}
the positivity conditions {\it i)-iii)}. However there can be many different solutions to these
conditions that may not correspond to the high-density liquid. In particular
the solution proposed by Samorodnitsky, Eq.~(\ref{gdelta}), is not suitable to
describe a liquid state, as we
do not expect to observe a large number of particles in contact (represented by a peak at $i=d$)
in the liquid phase. Thus we will first look for a function $g(i)$ verifying {\it i)-iii)},
not showing large peaks and departing continuously from the step function for
$\f\geq 0$.
We will show in the following that such a function exists up to
$\f = \f_{HNC}(\d) < \f_S(\d)$, see Fig.~\ref{fig:1}, and is indeed
given by the step function plus an exponentially small correction in $n$. 
We interpret this solution as describing
the liquid phase and show numerically that the solution of the HNC equations converges to this
solution for $n\to \io$.

Using the solution above we can compute the entropy of the liquid.
A crucial observation is that this entropy becomes negative for $\f \sim \log(n)/n$, \ie
very close to the VG bound. This means that the liquid phase must become unstable below this 
value of density, as the entropy of a discrete system must be positive. We then expect that
the system (\ref{GCpart}) will undergo a {\it phase transition} at a density
$\f \leq \log(n)/n$.

This behavior closely resembles the behavior of hard spheres in the continuum in the limit of
large space dimension. For this problem, we recently showed \cite{PZ06} that at a value of density close to 
the radius of convergence of the virial expansion (\ie to the VG value) the liquid phase becomes
unstable towards a glass phase where replica symmetry is broken.
In the glass phase the pressure rapidly increases and diverges at a maximum density (for the glass) 
which is found to be of the same order of the glass transition density.

By analogy with the problem in the continuum, we argue that also in this problem a glass
transition exists at a density $\f_K \leq \log(n)/n$ and that the glass phase should exist up
to a maximum density with the same scaling in $n$. This means that, {\it if no other phases
exist}, one should have $\f_c \sim \log(n)/n$, \ie the VG bound should be tight.
Unfortunately we are still not able to repeat the calculation of the equation of state of the
glass - that was done in \cite{PZ06} for the problem in the continuum - for this problem;
but we believe that the computation is feasible and we leave it for future work.

Clearly other phases
may exist at least for some special values of $n$, $d$. For instance,
we have some numerical evidence, for even $d$, 
of a first-order phase transition toward a phase in which
particles only occupy a subspace of the Hamming space. We will discuss this issue below.

In the following we will try to make these arguments more precise.

\subsection{The Fourier transform for $n\to\io$}

\begin{figure}[t]
\includegraphics[width=6cm]{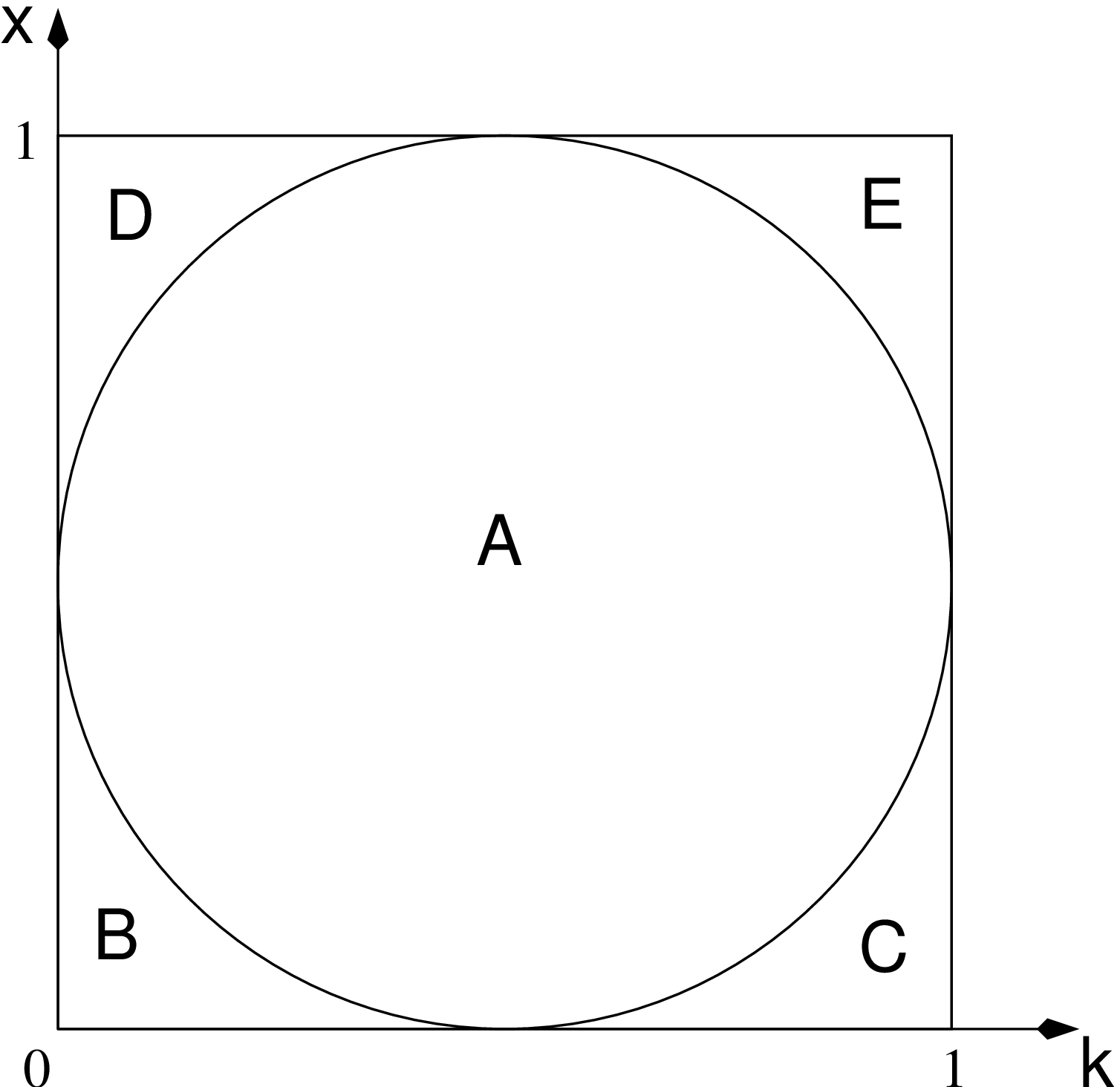}
\includegraphics[width=8cm]{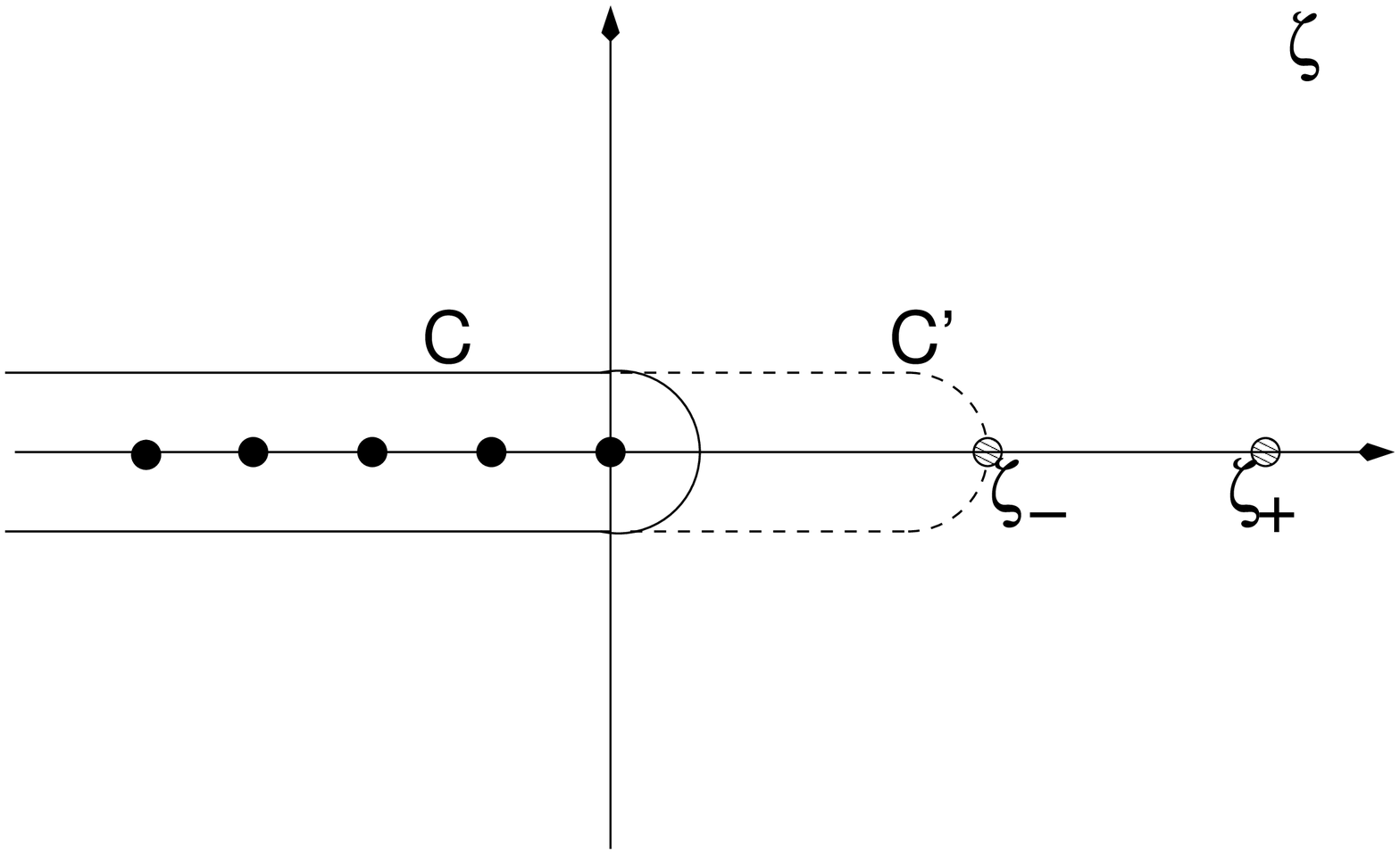}
\caption{
Left: the domain in the plane $(k,x)$ where the saddle points (\ref{SP}) are
complex is represented by region A.
Right: deformation of the integration contour to include the saddle point
$\z_-$ (this plot refers to region B, see text).
}
\label{fig:2}
\end{figure}

\begin{figure}[t]
\includegraphics[width=10cm]{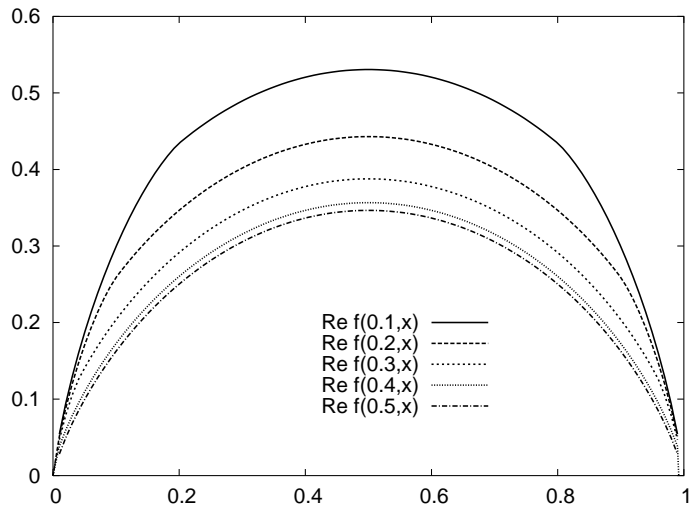}
\caption{
The function $\Re f(k,x)$ for different values of $k$ as a function of $x$;
from top to bottom $k=0.1,0.2,0.3,0.4,0.5$.
Recall that from Eq.~(\ref{S2}) $\Re f(k,x) = \Re f(1-k,x)$.
}
\label{fig:3}
\end{figure}

We begin by studying the properties of the Fourier transform of the delta function 
$g(j) = \d_{ij}$,
that is simply $\FF_n(a,i)$, for $n\to\io$. We define $k=a/n$ and $x=i/n$. We want
to compute the function $f(k,x)$ defined by
\beq
\Phi(k,x) = \FF_n(nk,nx) \sim \Re e^{n f(k,x)} \ , \hskip10pt n\to\io \ .
\eeq
Using the fact that the poles of the gamma function $\G(z)$ are in $z=-k$ with residual
$(-1)^k/k!$, we can rewrite Eq.~(\ref{FTsferico}) as
\beq
\FF_n(a,i) =\int_C dz \G(z) \frac{\G(a+1) \G(n-a+1)}{\G(a+z+1) \G(i+z+1) \G(n-a-i-z+1)} \ ,
\eeq
where the contour $C$ embraces the negative part of the real $z$ axis, see Fig.~\ref{fig:2}.
Using the Stirling formula $\G(n) \sim n^n e^{-n} \sqrt{2\p n}$, we get, changing the integration
variable to $\z=z/n$, and neglecting power-law prefactors,
\beq
\FF_n(nk,nx)=\int_C d\z \, e^{ n[\z\log\z - v(k) -(k+\z)\log (k+\z)
-(x+\z) \log (x+\z) - (1-k-x-\z) \log (1-k-x-\z)]} =
\int_C d\z \, e^{ n \s(\z)}
\ ,
\eeq
and we can evaluate the integral using the saddle point method. The saddle point equation is
\beq
\frac{\z(1-k-x-\z)}{ (k+\z) (x+\z)} = 1 \ ,
\eeq
with solutions
\beq\label{SP}
\z_\pm = \frac12 \left[ \frac12-k-x\right] \pm \frac12 
\sqrt{\left( k-\frac12\right)^2 + \left(x-\frac12\right)^2 -\frac14} \ .
\eeq
From the analysis of the position of the solutions in the complex plane one can deduce the following:
\begin{enumerate}
\item In the region $\left( k-\frac12\right)^2 + \left(x-\frac12\right)^2 -\frac14 \leq 0$
(region A in Fig.~\ref{fig:2})
the solutions are complex, and $\z_+ = \z_-^*$, $\s(\z_+) = [\s(\z_-)]^*$. Thus one has
\beq
\FF_n(nk,nx) \sim e^{n\s(\z_-)} + e^{n\s(\z_+)} = 
\Re e^{n\s(\z_-)} \sim e^{n \Re \s(\z_-)} \cos [n \Im \s(\z_-)] 
\hskip10pt \Rightarrow \hskip10pt f(k,x) = \s(\z_-) \ .
\eeq
\item In region B of Fig.~\ref{fig:2} the saddle points are real and positive with
$0 \leq \z_- \leq \z_+$, and $\s(\z_\pm)$ are also real with $\s(\z_+) > \s(\z_-)
> 0$.
The point $\z_-$ is then the closest saddle point to the original integration contour;
moreover, it is a local mimimum of $\s(\z)$ along the real axis, so it is
a maximum of $\s(\z)$ along the imaginary direction and the integration path can be deformed
to include it without crossing regions of $\z$ where $\Re\s(\z) > \s(\z_-)$, see Fig.~\ref{fig:2}. 
Then
\beq
\FF_n(nk,nx) \sim e^{n \s(\z_-)}\hskip10pt \Rightarrow \hskip10pt f(k,x) = \s(\z_-) \ .
\eeq
\item The behavior in the regions C,D,E can be obtained using the symmetries (\ref{S2}), (\ref{S3}),
(\ref{S4}). Alternatively one can always choose the closest saddle point to the integration path
that is a local minimum on the real axis:
it turns out that one has to choose $\z_-$ in the region E and $\z_+$ in the regions
C and D. With this choice the symmetries (\ref{S2}), (\ref{S3}),
(\ref{S4}) are respected.
\end{enumerate}
Finally one obtains
\beq
f(k,x) = \min_{\Re} [\s(\z_+),\s(\z_-)] \ ,  
\eeq
where $\min_\Re$ means that one has to take the solution with the smallest real part.
The real part of the resulting function $f(k,x)$ is an increasing function of $x$ for 
all $k$ and $x < 1/2$, see Fig.~\ref{fig:3}. 
This allows to compute the Fourier transform of the theta function
$\th(d-i)$ for $d < n/2$. Defining $f_x(k,x) = \partial_x f(k,x)$, we have
\beq\label{FourierTheta}
\begin{split}
\wh \th_\d(nk) &= \sum_{i=0}^n \FF_n(nk,i) \th(d-i) \sim
n \Re \int_0^\d dx \, e^{n f(k,x)} \sim 
n \Re \int_0^\d dx \, e^{n [f(k,\d) + f_x(k,\d) (x-\d) + \cdots]} \\
& \sim \Re \frac{e^{n f(k,\d)}}{f_x(k,\d)} \sim \Phi(k,\d) \ ,
\end{split}
\eeq
\ie the Fourier transform of the theta coincides with the one of the delta to
leading order in $n$ as long as $d < n/2$.
Similarly we can compute the Fourier transform of a function that vanishes outside a finite
interval and approaches zero linearly at the edge of the interval, \ie $h(i) =
\th(d-i) (d-i)$:
\beq\label{FourierLin}
\begin{split}
\wh h(nk) &\sim n^2 \Re \int_0^\d dx \, e^{n f(k,x)} (\d-x) \sim
n^2 \Re \int_0^\d dx \, e^{n [f(k,\d) + f_x(k,\d) (x-\d) + \cdots]} (\d-x) \\ &\sim
\Re e^{n f(k,\d)} \frac1{[f_x(k,\d)]^2} \sim \Phi(k,\d) \ ,
\end{split}\eeq
\ie to leading order in $n$ also this function is equal to $\Phi(k,\d)$.

\subsection{HNC for $n\to\io$}

\begin{figure}[t]
\includegraphics[width=10cm]{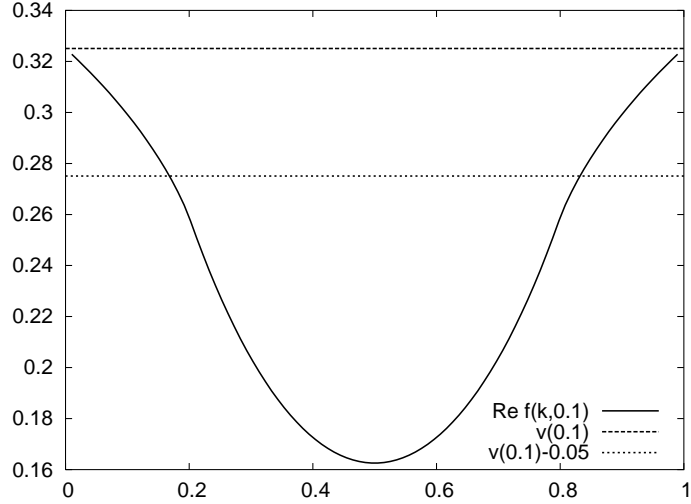}
\caption{
The function $\Re f(k,\d)$ as a function of $k$ for $\d=0.1$. The value of $v(\d)$ and
of $v(\d) - 0.05$ are also reported. The value of $k$ such that $\Re f(k,\d)$
intersects $v(\d)-\f$ is the value of $k_c(\d,\f)$, see Eq.~(\ref{eqkc2}).
}
\label{fig:4}
\end{figure}

We argue that for $n\to\io$ the solution to the HNC equation approaches a solution
of conditions {\it i)-iii)} which shows no large peaks. In particular we will look
for a solution of the form $g(i) = \th(i-d+1) [1+\exp(n \wt h(i))]$, $\wt h(i)
\leq 0$, \ie a solution
differing from the step function (that describes the liquid for $\f < 0$)
by an exponentially small quantity.
Conditions {\it i)-iii)} are, for $x = i/n$ and $k=a/n$, and recalling that $h(i)=g(i)-1$,
\beq\label{posy}
\begin{cases} \r \wh h(k) \geq -1 \ , \\
h(x) \geq -1 \ , \\
h(x) = -1 \ , \hskip30pt x \leq \d \ .
\end{cases}
\eeq
First we will check that for $\f < 0$ the step function $g(x) = \th(x-\d)$, \ie
$h(x) = -\th(\d-x)$, satisfies the conditions above. 
Using Eq.s~(\ref{fdef}) and (\ref{FourierTheta}),
\beq
\r \wh h(k) = -\r \Phi(k,\d) \geq -1
\hskip10pt \Leftrightarrow \hskip10pt \Re e^{n[\f - v(\d) + f(k,\d)]} = 
e^{n[\f - v(\d) + \Re f(k,\d)]} \cos[n\Im f(k,\d)] \leq 1 \ ,
\eeq
and the latter relation is equivalent to
\beq\label{condth}
\Re f(k,\d) \leq v(\d)-\f \ .
\eeq
In Fig.~\ref{fig:4} the function $\Re f(k,\d)$ is reported as a function of $k$ for
a representative value of $\d$. It assumes its maximum in $k=0$ and $k=1$ and
$f(0,\d) = v(\d)$. Thus the inequality (\ref{condth}) is always satisfied if
$\f \leq 0$,
so that in this region (which is also the region where the virial series converges)
we argue that $h(x) = -\th(\d-x)$ describes the liquid phase for $n\to\io$. This can
also be checked by a direct evaluation of the leading terms in the (convergent) virial series, 
see~\cite{FP99}. Following \cite{FP99} we also argue that the HNC resummation contains
all the relevant diagrams for $\f < 0$, so we can use it to obtain the free energy of the
liquid.
Substituting the result for $h(x)$ in Eq.~(\ref{HNCFree2}) 
the last term is exponentially small in $n$ and one obtains,
up to exponentially small corrections,
\beq\label{HNCfneg}
\begin{split}
&\Psi(\r) = -S(\r) = \frac{\r V_d}2 + \ln \r - 1 \ , \\
&\wt P \equiv \frac{P}{\r} = \r \frac{d\Psi}{d\r} = 1 + \frac{\wt \r}{2} \ ,
\end{split}\eeq
where $\wt P$ is the reduced pressure. 
As found in \cite{FP99} we find that the entropy is given by the ideal gas term
plus the first virial correction.
Note that $\wt \r$ is exponentially small
for $\f <0$ so the system behaves essentially as an ideal gas.

For $\f > 0$ the function $h(x)$ cannot be
given by $-\th(\d-x)$ as this function does not respect the positivity conditions
(\ref{posy}). We follow the strategy of \cite{PS00} and decompose
$h(x) = h_0(x) + h_1(x)$ assuming that $h_0(x)$ vanishes for $x>\d$ and $h_1(x)$ is a
continuous function of $x$. We call $Y \equiv h_1(\d) = -1-h_0(\d^-)$,
\ie $h_0(\d^-) = -(1+Y)$.
As $h_0(x)$ vanishes for $x>\d$, its Fourier transform is given by (\ref{FourierTheta}):
\beq
\wh h_0(k) = -(1+Y) \Phi(k,\d) \ ,
\eeq
and the condition $\r\wh h(k) = \r\wh h_0(k) + \r \wh h_1(k) \geq -1$ becomes
\beq
\wh h_1(k) \geq -\frac1\r +(1+Y)\Phi(k,\d) \ .
\eeq
We choose the simplest solution to the previous equation,
\beq\label{h1k}
\wh h_1(k) = \begin{cases}
-\frac1\r +(1+Y)\Phi(k,\d)   \hskip30pt \text{for  } -\frac1\r +(1+Y)\Phi(k,\d) \geq 0 \ , \\
0 \hskip114pt   \text{for  } -\frac1\r +(1+Y)\Phi(k,\d) < 0 \ ,
\end{cases}
\eeq
and we will show that it gives, in real space, 
an exponentially small correction with respect to the step function.
The function $\wh h_1(k)$ has a two singularities when 
\beq\label{eqkc1}
(1+Y) \r \Phi(k,\d) = 1 
\hskip10pt \Leftrightarrow \hskip10pt
\f - v(\d) + f(k,\d) + \frac1n \ln(1+Y) = 0
\ ,
\eeq
but the above equation is well defined only if its solutions lie in the region where 
$f(k,\d)$ is real, otherwise the function $\Phi(k,\d)$ will oscillate very fast.
This will impose some restrictions to the values of $\f$ and $d$. 

First we will restrict to 
odd $d$ in order to have the symmetry $\Phi(1-k,\d) = \Phi(k,\d)$; for even $d$
we have the opposite symmetry $\Phi(1-k,\d) = -\Phi(k,\d)$ and $f(k,\d)$ has an
imaginary part for $k > 1/2$.
This follows from $\Phi(k,\d)=\FF_n(a,d)$ (with $a=nk$ and $d=n\d$) and from Eq.~(\ref{S2}).
For odd $d$ Eq.~(\ref{eqkc1}) will have two solutions $k_c$, $1-k_c$ due to the symmetry, see Fig.~\ref{fig:4}.
Note that the opposite restriction
was applied in the numerical computation of \cite{BJ01} where only the case of even $d$ has
been considered.

Next, we look for a solution $k_c$ outside the region A of Fig.~\ref{fig:2},
as in region A we already know that $f(k,\d)$ is not real. This means that the
maximum possible value for $k_c$ is
\beq
k_{max}(\d)=\frac12-\sqrt{\frac14-\left(\d-\frac12\right)^2} = \frac12 - \sqrt{\d(1-\d)} \ ,
\eeq
that is the boundary of the region where $f(k,\d)$ is real, see Fig.~\ref{fig:2}. 

As we will self-consistently verify at the end, under the restrictions above
one has $Y \sim e^{-n \kappa}$, then Eq.~(\ref{eqkc1}) becomes
\beq\label{eqkc2}
f(k,\d) = v(\d) - \f \ .
\eeq
For $\f \leq 0$ it has no solutions as discussed above, so $\wh h_1(k) = 0$, then
$h_1(x)=0$, $Y=0$ and we recover the step function solution.
For $\f > 0$, $k_c$ increases from $0$ to $k_{max}(\d)$ and reaches the
boundary of region A, using Eq.~(\ref{eqkc2}), exactly at
\beq\label{fHNC}
\f = v(\d)-f(k_{max}(\d),\d) = \f_{MRRW}(\d)/2 = \f_S(\d) \ ,
\eeq
where $\f_{MRRW}(\d)$ has been defined in Eq.~(\ref{MRRW}).

We have now to compute $h_1(x)$ from $\wh h_1(k)$.
The function $\wh h_1(k)$ verifies $\wh h_1(k)=\wh h_1(1-k)$ and 
is nonzero only for $k<k_c$ and $k>1-k_c$; it vanishes linearly close
to $k_c$, $\wh h_1(k) \sim (1+Y) \Phi_k(k_c,\d) (k-k_c)$, and 
$\Phi_k(k,\d) = \partial_k \Phi(k,\d)$ is real in $k=k_c$.
If we restrict to even $i$, we have also $\Phi(x,k)=\Phi(x,1-k)$ by
symmetry (\ref{S3}). We can do that because it is possible to show that, in
the case of odd $d$ that we are considering, one has $h(i-1)=h(i)$ for even
$i$; thus it is enough to compute $h(i)$ for even $i$ 
(see the next section, Appendix~\ref{app:A} and Fig.~\ref{fig:6} 
for a detailed discussion of this tricky point).
Using Eq.~(\ref{FourierLin}) we have then
\beq\label{h1x}
\begin{split}
h_1(x) &= \frac{2n}{2^n} \int_0^{1/2} dk \, \Phi(x,k) \wh h_1(k) \sim
\frac{2n}{2^n} (1+Y) e^{n f(k_c,\d)} n f_k(k_c,\d) \, \Re \int_0^{k_c} dk \, e^{n f(x,k)} (k-k_c) \\
&\sim \frac{1+Y}{2^{n-1}} e^{n f(k_c,\d)} \, \Re e^{n f(x,k_c)} 
\frac{f_k(k_c,\d)}{[f_x(x,k_c)]^2} \ .
\end{split}
\eeq
Keeping only the leading terms (exponentials of $n$) we obtain the self consistency
equation for $Y$:
\beq
Y = h_1(\d) = (1+Y) e^{n [ f(k_c,\d) + f(\d,k_c) -\ln 2 ]} \ .
\eeq
If $k_c < k_{max}$ we have $ f(k_c,\d) + f(\d,k_c) -\ln 2 < 0$, so that
$Y$ is exponentially small and is given by
\beq
Y = e^{n [ f(k_c,\d) + f(\d,k_c) -\ln 2 ]} = 
e^{n [v(\d)+v(k_c)-2\f -\ln 2]} \ .
\eeq
where we used the relation $f(x,k) = f(k,x) + v(k) - v(x)$ that follows from Eq.~(\ref{S1}) and
Eq.~(\ref{eqkc2}).
Finally, the function $\wh h(k)=\wh h_0(k) + \wh h_1(k)$ is given, using 
Eq.~(\ref{h1k}) and $Y \ll 1$, by
\beq\label{hkHNC}
\r \wh h(k) = \begin{cases}
-1   \hskip80pt \text{for  } k < k_c \text{ and } k > 1-k_c \ , \\
-\r \Phi(k,\d)   \hskip50pt   \text{for  } k_c < k < 1-k_c \ .
\end{cases}
\eeq
We can rewrite Eq.~(\ref{h1x}) using $Y \ll 1$, neglecting non-exponential
prefactors, and recalling that $f(k_c,\d)=v(\d)-\f$, as
\beq
h_1(x) = \frac{e^{n[v(\d)-\f]}}{2^n} \Phi(x,k_c) = \frac{1}{\r}
\frac{1}{2^n}\Phi(x,k_c) \ ;
\eeq
note that from Eq.~(\ref{FTsferico}) it follows that $2^{-n} \Phi(x,k_c)$ is
the Fourier transform of $\d(k-k_c)$.
Finally, $h(x)$ is given, from Eq.~(\ref{h1x}), by
\beq\label{hx}
\begin{split}
&h(x) = \begin{cases}
-1 \hskip35pt x< \d \ , \\
\Re e^{n \wt h(x)}  \hskip22pt x \geq \d \ , 
\end{cases} \\
&\wt h(x) = v(\d)-\f + f(x,k_c) -\ln 2  \ .
\end{split}
\eeq
The solution above is defined up to the value $\f_{S}$ of the density given
by Eq.~(\ref{fHNC}).
Indeed, the solution $k_c$ of Eq.~(\ref{eqkc1}) is given by Eq.~(\ref{eqkc2}) only if it is
in the region where $f(k,\d)$ is real. Otherwise, oscillations are present in $\Phi(k,x)$ and
the solution is not well defined. Moreover, if $Y$ is not exponentially small again the
solution above fails. Both these conditions seem to be violated for $\f > \f_{S}(\d)$.

There is however another condition to be imposed, namely that $\wt h(x) \leq
0$; otherwise the solution will be exponentially large and again we do not
expect that for a liquid phase. The maximum of $\wt h(x)$ is attained in
$x=0$, $x=1$, and is given by $v(\d)+v(k_c)-\f-\ln 2$. The condition $\wt h(x) \leq
0$ then requires
\beq
v(\d)+v(k_c)-\f-\ln 2 \leq 0 \ .
\eeq
A numerical solution of the previous equation (recall that $k_c$ depends
implicitly on $\f$) gives the stability threshold $\f_{HNC}(\d)$, which is
reported in Fig.~\ref{fig:1}.
For $\f \geq \f_{HNC}$ our solution starts to exhibit diverging oscillations
at large $x$. This is observed also in the numerical solution of the HNC
equations, see below.
Above $\f_S(\d)$ either the solution does not exist anymore or it yields a value of $Y$ that
is exponentially diverging with $n$. In both cases the solution does not describe a liquid
phase.
We will see however
that we are not really interested in so high values of the density as 
the liquid phase becomes unstable at a
much lower density.

\subsection{HNC entropy}

The HNC free energy is the canonical free energy, that for hard spheres is simply
$-S(\r)$. For $\f \leq 0$ it is given by Eq.~(\ref{HNCfneg}), and we showed
that for $\f > 0$ only exponentially small corrections appear. 
Neglecting these corrections we have from Eq.~(\ref{HNCfneg})
\beq
S(\r) = 1 - \log \r - \frac12 \r V_d \sim 1 - n \f + nv(\d) - \frac12 \log [2\p\d(1-\d)] -
\frac12 \log n - \frac12 e^{n\f} \ ,
\eeq
in the full range $\f < \f_{HNC}(\d)$.
An interesting observation is that $S(\r)$ becomes negative at a value
of density given by, keeping only the leading terms,
\beq
S(\f) \sim -n \f +n v(\d) - \frac12 \log n - \frac12 e^{n\f} = 0 \ ,
\eeq
and the solution is, at first order in $n\to \io$,
\beq
\f_0 = \frac1n \log [ 2 n v(\d) -3 \log n ] \ .
\eeq
As for a discrete system $S(\f) \geq 0$, 
this means that the liquid phase must become unstable at a density $\f < \f_0
\sim \log(n)/n$.

\subsection{Numerical solution of the HNC equations}

\begin{figure}[t]
\includegraphics[width=8cm]{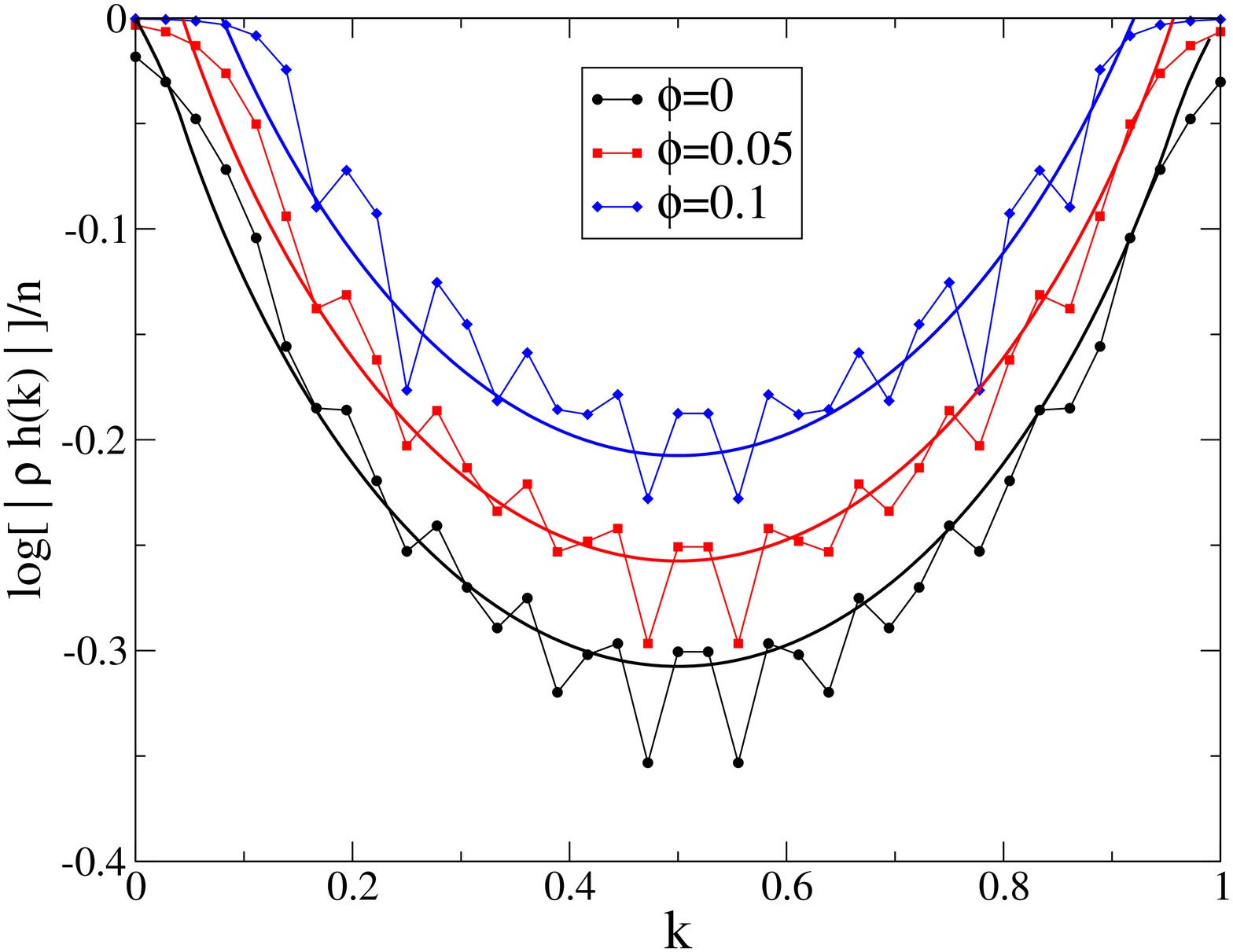}
\includegraphics[width=8cm]{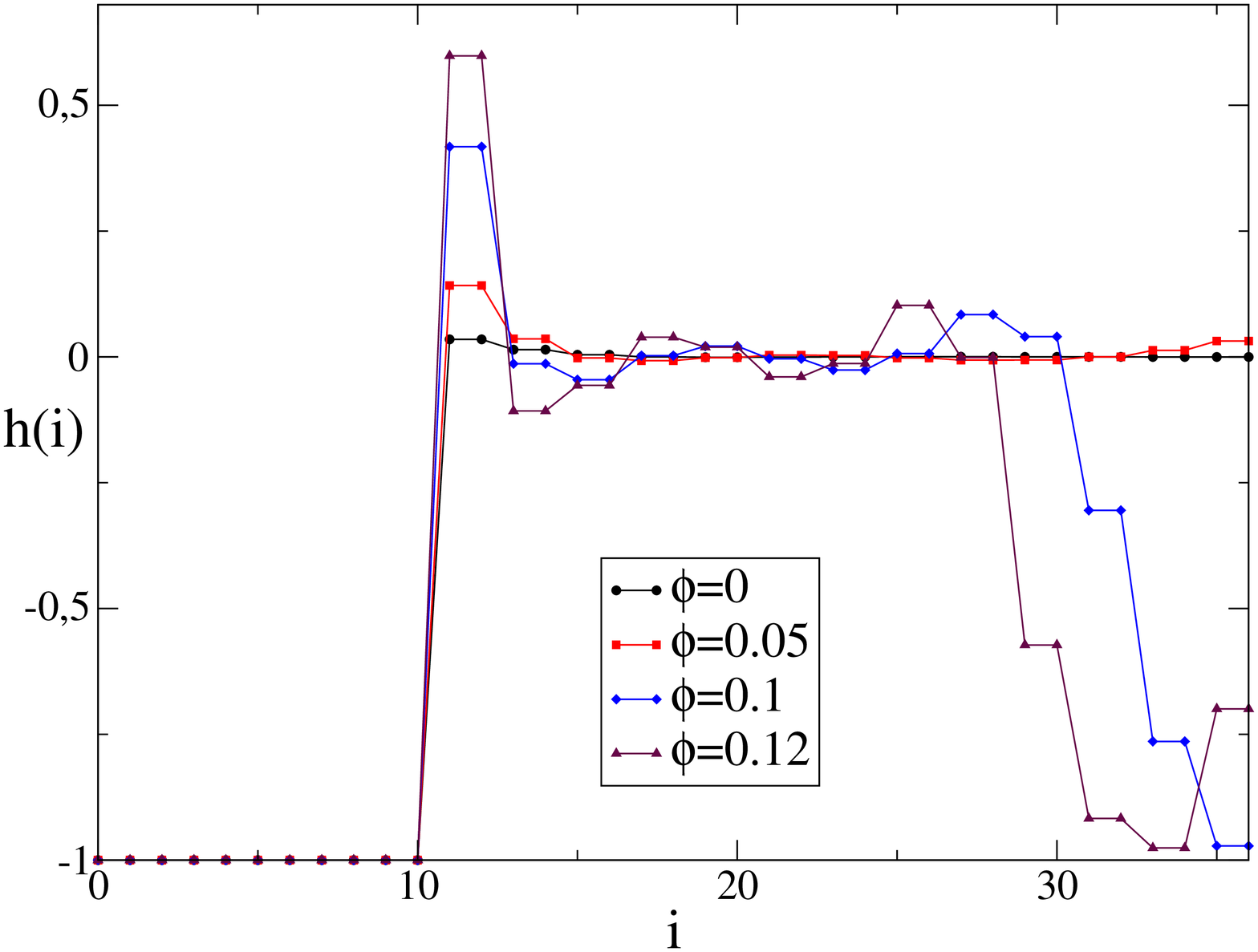}
\caption{
Comparison of the numerical solution of the HNC equations with the asymptotic
solution of the positivity conditions.
Left: The function $\ln[|\r \wh h(k)|]/n$ obtained 
numerically is reported
as a function of $k$ for $n=36$ and $d=11$ (\ie $\d=0.305$) and $\f=0,0.05,0.1$. 
The lines are obtained from
Eq.~(\ref{numericeq}).
Right: The function $h(i)$ for $n=36$, $d=11$ and $\f=0,0.05,0.1,0.12$. 
At large $\f$ we start to observe diverging oscillations at high
values of $i$, see text.
}
\label{fig:5}
\end{figure}

We will now compare the asymptotic solution with a numerical solution of the HNC equations 
(\ref{HNCeq}). The latter are solved using a standard iterative algorithm.
For graphical convenience we report the function $\frac1n \log | \r \wh
h(k)|$, that from Eq.~(\ref{hkHNC}) is given by
\beq\label{numericeq}
\lim_{n\to\io} \frac1n \log|\r\wh h(k)| = \begin{cases}
0  \hskip220pt \text{for  } k < k_c \text{ and } k > 1-k_c \ , \\
\Re f(k,\d)+\f-v(\d) + \frac1n \log | \cos n \Im f(k,\d)|   \hskip30pt   \text{for  } k_c < k < 1-k_c \ .
\end{cases}
\eeq
The last term is nonzero only for $k \in [k_{max}(\d),1-k_{max}(\d)]$ and
gives rise to oscillations whose frequency increases on increasing
$n$. Moreover at the values of $k$ where $\cos n \Im f(k,\d)=0$ the last term
diverges. These values accumulate in the interval $k \in
[k_{max}(\d),1-k_{max}(\d)]$
for large $n$.
However, for the relativaly small values of $n$ considered here, the are no
integer values of $a=nk$ where $\cos n \Im f(k,\d)=0$ and we can compare the
function  $\frac1n \log | \r \wh
h(k)|$ with its asymptotic limit neglecting the last term.
Moreover we can compare $h(i)$ with the analytical expression (\ref{hx}). 
The comparisons are encouraging as shown in Fig.~\ref{fig:5}. 
We find that $\wh h(k)$ agrees well with Eq.~(\ref{hkHNC}) and,
for small $\f$, $h(i)$ is essentially the step function plus
a small corrrection. On increasing the density large oscillations appear at
large $i$, as predicted by Eq.~(\ref{hx}). Unfortunately a quantitative comparison of $h(i)$ with the
asymptotic expression requires either the evaluation of finite $n$ corrections,
due to the small values of $n$ we can investigate numerically, or
the (difficult) investigation of much larger values of $n$, see \eg
\cite{BJ01}.

\subsection{Is there a glass transition?}

The fact that the entropy becomes negative seems to indicate the existence of
a phase transition. By analogy with the continuum problem, where we showed
that a glass transition happens at a similar value of the density \cite{PZ06}, we can
conjecture that a glass transition will happen also in this problem.

If this is the case, one can show by general arguments and by analogy with the
continuum problem that $S_{glass}(\f) \leq S_{liquid}(\f)$ (this is
because a downward jump of the compressibility is expected at the glass
transition \cite{PZ05}). This means that the entropy of the glass will vanish
at a density $\f < \f_0 \sim \log(n)/n$. Then both the liquid and the glass
phase will disappear for $\f \sim \log(n)/n$. If there are no other phases
(such as a ``crystal''), this scenario indicates that the Varshamov-Gilbert 
bound should give asymptotically the exact result.

Note also that the correlation function used by Samorodnitsky,
Eq.~(\ref{gdelta}) is similar to the one we expect for the glass phase
(see the discussion in \cite{PZ05,PZ06}). This means that, if the picture
above is correct, the $g(i)$ in Eq.~(\ref{gdelta}) should correspond to realizable 
packings only up to a density $\f \sim \log(n)/n$. 
We hope that further work will clarify this issue, see also the discussion in
\cite{To06,Le06}.

\subsection{Other instabilities}
\label{sec:otherinst}

\begin{figure}[t]
\includegraphics[width=8cm]{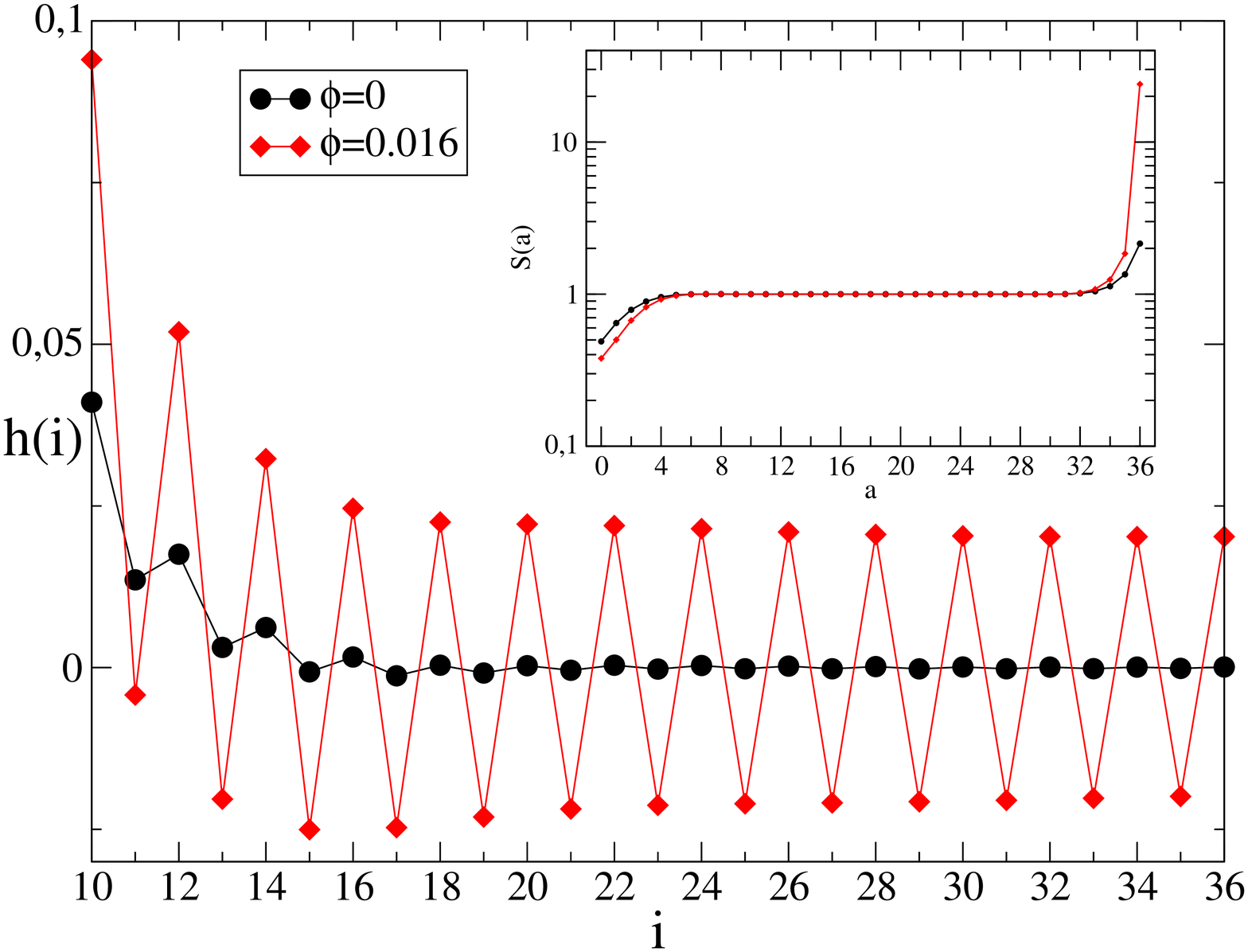}
\includegraphics[width=8cm]{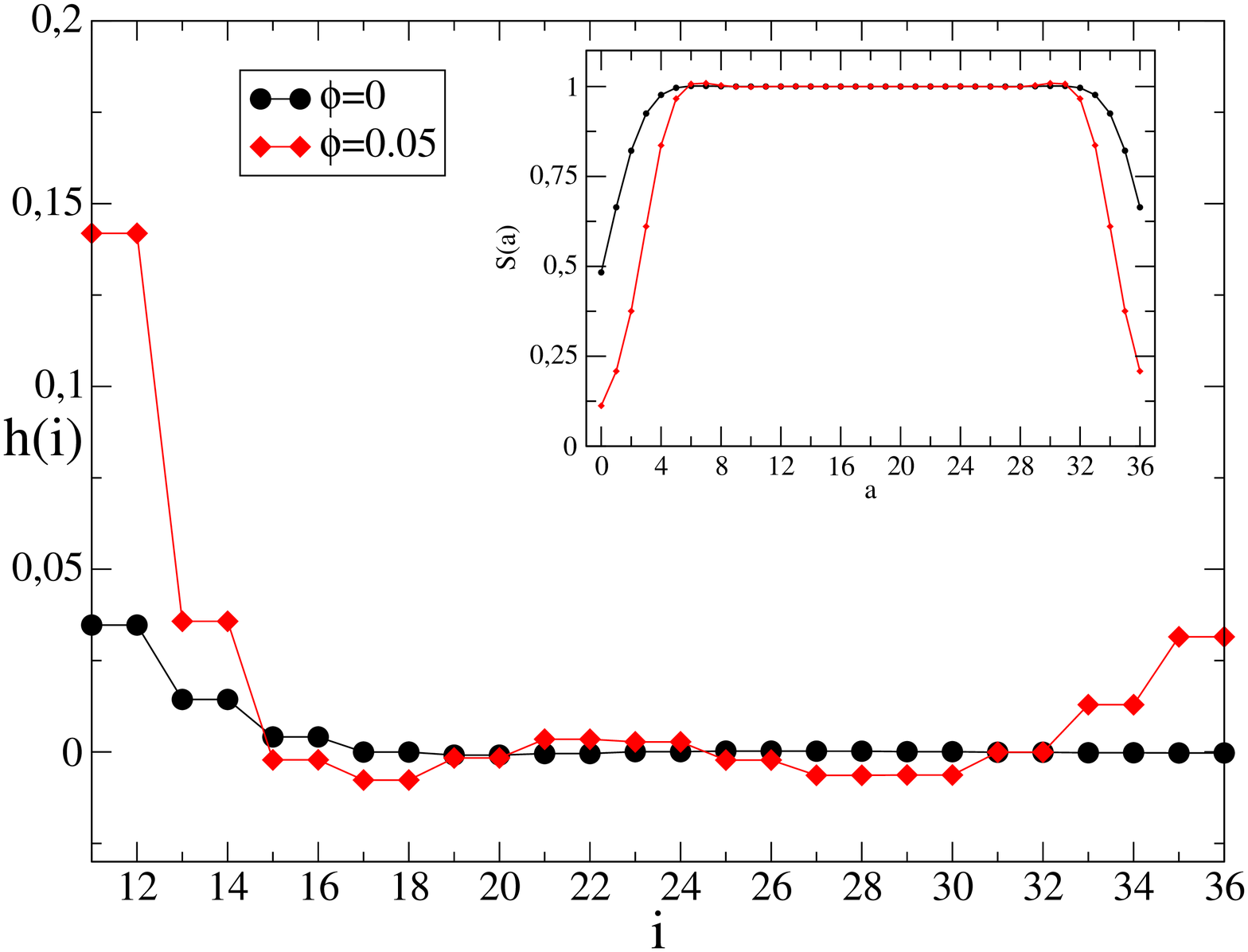}
\caption{Left:
correlation function $h(i)$ for $n=36$, $d=10$ and $\f=0,0.016$.
Inset: structure factor $S(a)=1+\r \wh h(a)$ for the same values
of the parameteres.
On increasing $\f$ above $\f=0$ strong oscillations appear in
$h(i)$, related to a strong peak in $S(a)$ for $a=36$.
Right: same plot for $n=36$, $d=11$ and $\f=0,0.05$. In this case
the oscillations are absent, $S(a)$ has a different behavior for large
$a$, and $h(i)$ decays in steps of two. The different (approximate) symmetry
of $S(a)$ is related to the symmetry (\ref{S2}).
}
\label{fig:6}
\end{figure}

We would like now to consider a different instability of the liquid that is
observed for even $d$. Indeed, $d$ is the smallest possible distance between
any pair of spheres. 
Let us split the Hamming space $H^n_2$ into two spaces $H^n_2 = H^n_e \cup
H^n_o$, where $H^n_e \equiv \{ s \in H^n_2 | d(s,0) \text{ is even} \}$
and $H^n_o \equiv \{ s \in H^n_2 | d(s,0) \text{ is odd} \}$.
Recall that $\r g(i)=\r [1+h(i)]$ is the probability of finding a particle in
a given point at distance $i$ from the origin,
given that there is a particle in the origin.

At high density,
given that there is a sphere $s_0$ in the origin, it will be convenient, say, to place
the second sphere $s_1$ at the minimum distance $d$, thus $s_1 \in
H^n_e$. Then we can put another sphere $s_2$ at minimum distance $d$ from
$s_1$, so that $s_2 \in H^n_e$, and so on. It is clear that this picture is
oversimplified but still we can expect that at large enough density particles
will concentrate on one of the subsets $H^n_{e,o}$, depending whether there
is or not a particle in the origin.

This is indeed what we
found in the solution of the HNC equations. The partial localization of the
particles on a subset $H^n_{e,o}$ is revealed by oscillations in $h(i)$, which
is much bigger on the even values of $i$, see Fig.~\ref{fig:6}.
The oscillations in $h(i)$ are related to a strong peak developping in 
$S(a)=1 + \r \wh h(a)$ for $a=n$ on increasing $\f$ above $0$. Due to this
growing peak, the HNC equations rapidly become unstable for small values of
$\f$ (much smaller than for odd $d$). Preliminary Montecarlo simulations seems
to indicate the existence of a first order phase transition to a phase in
which particles are completely localized on a sublattice at a critical value
of $\f$. 

For odd $d$ a related phenomenon occurs. Indeed, by looking
at the correlation function $h(i)$ for odd $d$ (see right panel in
Fig.~\ref{fig:6}), we see that we have $h(i)=h(i+1)$ for odd $i$, \ie that
the correlation function decays in steps of two. This can be understood by the
following argument. Consider a system of sequences $s' \equiv (s,s_{n+1}) \in
H^{n+1}_e$, \ie with the constraint that $d'(s',0)$ is even, 
where $d'(s',0)=d(s,0)+s_{n+1}$
is the distance in the Hamming space $H^{n+1}_2$.
This constraint fixes the last
bit $s_{n+1}$: as $s_{n+1}=0$ if $d(s,0)$ is even and $s_{n+1}=1$ if
$d(s,0)$ is odd. 
If we consider in the original space $H^n_2$
a sphere at distance $d(s,0)=i$ with $i$ odd we have $d'(s',0) = i+1$, but also
if $d(s,0)=i+1$ we have $d'(s',0)=i+1$. Thus, spheres at distance $i$
and $i+1$ with $i$ odd have the same distance $i+1$ (even) from the origin in $H^{n+1}_2$.

This means that the original problem on $H^n_2$ with minimum distance $d$ odd
can be mapped into a problem on $H^{n+1}_e$ with $d' = d+1$ even.
However, in the new
problem, {\it particles that were at a distance $i$ and $i+1$ from the (old) origin 
(for odd $i$) are at the same distance from the (new) origin}, thus the
probability to have a particle at distance $i$ (odd) and $i+1$ in the
original problem, given that there is a particle in the origin, should be the same.

Note that the argument cannot be repeated for even $d$ as in this case we
should map the problem into a problem on $H^{n+1}_e$ with distance $d'$ odd,
and this is inconsistent as discussed above.

\section{Conclusions}
\label{sec:conclusions}

We discussed the high-density behavior of a system of hard spheres on the
hypercube $H^n_2$ from a physical point of view, trying to understand the mechanisms
that determine the maximum density of the system.

First we found a possible asymptotic solution for the liquid correlation in
the limit $n\to\io$, and we showed that this solution yields a negative
entropy for $\f \sim \log(n)/n$, \ie for a number of particles 
$N \sim \frac{2^n}{V_d} n^{1/n}$, very close to the Varshamov-Gilbert lower
bound. 

On this ground we argue that a phase transition must exist towards a different
phase. The nature of this transition is still unknown, but we presented some
arguments in favor of a glass transition (basically, the analogy with the
problem of hard spheres in $\RRR^n$ for $n\to\io$ \cite{PZ06}) and for a first
order transition toward a phase in which the particles are constrained to a sublattice
$H^n_{o,e}$ for even $d$ (some insight from the
solutions of the HNC equations and from preliminary Montecarlo simulations).

Both these possibilities require further investigation. In particular, the
study of the glass transition requires the computation of the replicated
partition function of the system following \cite{PZ05,PZ06}. This is more
difficult in this discrete problem as we cannot make a Gaussian {\it ansatz}
for the single particle density. The study of the first order transition will
require more extensive numerical simulations.

Moreover, there is also the possibility that, at least for some particular values of 
$n$ and $d$, some particular ``crystal-like'' configurations of high density exist.
Unfortunately, our approach is based on a low density expansion (the virial series) so
it is unable to capture the existence of such special configurations.

The results presented here are not conclusive but in our opinion may lead to
new ideas on how to improve the current bounds on $R(\d)$. We hope, in particular, that
future work will clarify whether a glass transition exist or not in this system.

\acknowledgments

We thank B.~Scoppola for pointing out the problem and for his
comments on this work. 
F.Z. is supported by the EU Research Training Network 
STIPCO (HPRN-CT-2002-00319), and wishes to thank G.~Biroli, A.~Giuliani,
J.~L.~Lebowitz, A.~Montanari, R.~Monasson, S.~Torquato for many useful
discussions.

\appendix
\section{More on symmetries}
\label{app:A}

We will discuss here the reason why we must restrict to even values of $i$
in computing $h_1(x)$, see Eq.~(\ref{h1x}). As we discussed in section \ref{sec:otherinst}
for odd $d$ the function $h(i)$ has the property that $h(i-1)=h(i)$ for even
$i$. This is a consequence of a hidden symmetry, namely the possibility to map
the problem in $H^n_2$ in a problem in $H^{n+1}_e$.

It can be shown that if $h(i)$ has such a symmetry, its Fourier transform $\wh
h(a)$ has the symmetry $\wh h(a) = \wh h(n-a+1)$, $a=1,\cdots,n$, due to the
structure of the matrix $\FF_n(a,i)$. This is consistent with the observation
that in the limit $n\to\io$ with $a=nk$, $\wh h(k)=\wh h(1-k)$, as comes out
from Eq.~(\ref{hkHNC}). However we are discarding a factor $1/n$. 

When inverting
the Fourier transform to recover $h(i)$, we find that if $i$ is even we get a
meaningful result, while if $i$ is odd, the symmetry $\wh h(k)=\wh h(1-k)$, if
interpreted as $\wh h(a)=\wh h(n-a)$
leads simply to $h(i)=0$ due to the symmetry (\ref{S2}). The direct
computation of $h(i)$ for odd $i$ would require a more refined calculation,
however it is much simpler to compute $h(i)$ for even $i$ and use the identity $h(i-1)=h(i)$.

This procedure seems to produce meaningful results as evidenced by the
positive agreement with numerical data, see Fig.~\ref{fig:5}.


\begin{thebibliography}{99}

\bibitem{VG} E.~N.~Gilbert, {\it A comparison of signalling alphabets}, Bell Syst. Tech. Jnl.
{\bf 31}, 504--522 (1952); R.~R.~Varshamov, {\it Estimate of the number of signals in error correcting
codes}, Dokl. Akad. Nauk SSSR {\bf 117}, 739--741 (1957).

\bibitem{De73}
P.~Delsarte, {\it An algebraic approach to the association schemes of coding theory}, 
Philips Res.~Rep.~Suppl. {\bf 10}, 1--97 (1973).

\bibitem{MRRW77} R.~J.~McEliece, E.~R.~Rodemich, H.~Rumsey and L.~R.~Welch,
{\it New upper bounds on the rate of a code via the Delsarte-MacWilliams inequalities},
IEEE Trans.Inform.Theory {\bf 23}, 157--166 (1977).

\bibitem{PS99} A.~Procacci and B.~Scoppola, {\it Statistical mechanics approach
to coding theory}, J.~Stat.~Phys. {\bf 96}, 907--912 (1999).

\bibitem{BJ01}
A.~Barg and D.~B.~Jaffe, {\it Numerical results on the asymptotic rate of binary codes},
in ``Codes and Association Schemes'' (A. Barg and S. Litsyn, Eds.), Amer.~Math.~Soc.,
Providence, 2001

\bibitem{Sa01}
A.~Samorodnitsky, {\it On the Optimum of Delsarte's Linear Program},
Journal of Combinatorial Theory, Series A {\bf 96}, 261--287 (2001).

\bibitem{FP99} H.~L.~Frisch and J.~K.~Percus,
{\it High dimensionality as an organizing device for classical fluids},
Phys.~Rev.~E {\bf 60}, 2942--2948 (1999).

\bibitem{PS00} G.~Parisi and F.~Slanina, {\it A toy model for the mean-field theory
of hard-sphere liquids}, Phys.~Rev.~E {\bf 62}, 6554--6559 (2000).

\bibitem{PZ06} G.~Parisi and F.~Zamponi, {\it Amorphous packings of hard spheres in
large space dimension}, cond-mat/0601573.

\bibitem{To06} O.~U.~Uche, F.~H.~Stillinger and S.~Torquato, {\it On the
    realizability of pair correlation functions}, Physica A {\bf 360}, 21--36 (2006);
S.~Torquato and F.~H.~Stillinger, {\it New provisional lower bounds on the
    optimal density of sphere packings}, submitted.

\bibitem{Hansen} J.-P.~Hansen and I.~R.~MacDonald, {\it Theory of simple
    liquids} (Academic Press, London, 1986).

\bibitem{kraw} Eric W. Weisstein, {\it Krawtchouk Polynomial}, 
From MathWorld--A Wolfram Web Resource, http://mathworld.wolfram.com/KrawtchoukPolynomial.html

\bibitem{Le75} A.~Lenard, {\it States of classical statistical mechanics of
    infinitely many particles: I, II}, Arch.~Rational.~Mech.~Anal. {\bf 59},
    219--239, 241--256 (1975). 

\bibitem{Le06}
T.~Kuna, J.~L.~Lebowitz and E.~Speer, {\it On the realizability of point
  processes with specified one and two particle densities}, in
Math.Frosch.Oberwolfach {\bf 43} (2004), Large Scale Stochastic Dynamics,
ed. C.~Landim, S.~Olla, and H.~Spohn; O.~Costin and J.~L.~Lebowitz, {\it On
  the construction of particle distributions with specified single and pair
  densities}, J.~Phys.~Chem. {\bf 108}, 19614 (2004).

\bibitem{PZ05} G.~Parisi and F.~Zamponi, {\it The ideal glass transition of
    hard spheres}, J.~Chem.~Phys. {\bf 123}, 144501 (2005).

\end{thebibliography}
\end{document}